\documentclass[prc,aps,showpacs]{revtex4}
\usepackage{graphicx}
\begin{document}
\title
{Production of $^{93,94,95,96}$Tc through $^{7}$Li+$^{nat}$Zr and $^{9}$Be+$^{nat}$Y reactions: Measurement of excitation functions}
\author{Moumita Maiti}
\affiliation{Chemical Sciences Division, Saha Institute of Nuclear Physics, 1/AF, Bidhannagar, Kolkata-700064, India.}
\author{Susanta Lahiri}
\affiliation{Chemical Sciences Division, Saha Institute of Nuclear Physics, 1/AF, Bidhannagar, Kolkata-700064, India.}
\begin{abstract}
For the first time two separate production routes of Tc radionuclides have been studied bombarding $^{7}$Li on $^{nat}$Zr and $^{9}$Be on $^{89}$Y. Excitation functions of the evaporation residues produced in those reactions have been measured using stacked-foil technique followed by the $\gamma$-spectrometric studies in the energy range 37-45 MeV and 30-48 MeV respectively. Measured excitation functions have been compared with those calculated using the nuclear reaction model codes PACE-II and ALICE91. Experimental results show good agreement with the theoretical predictions. Compound nuclear reaction is the key mechanism in producing evaporation residues.

\end{abstract}
\maketitle

\section{Introduction}

Tc is the element of interest since last three decades because of its several elegant features and practical applications e.g., in the field of nuclear. Out of identified isotopes of Tc, 21 are proton rich, having $\beta^{+}$ and/or $\epsilon$ decay mode, while rest 20 are neutron rich, mostly $\beta^{-}$ emitter. Neutron rich Tc isotopes are short-lived except $^{98}$Tc (T$_{1/2}$=4.2 My) and $^{99}$Tc (T$_{1/2}$=2.111$\times 10^{5}$ y). The only odd-even radioisomer, $^{99m}$Tc is extensively used in nuclear medicine because of its half-life (T$_{1/2}$=6.01 h) and 140.474 keV $\gamma$-ray, both suitable for {\it in vivo} imaging. In fact, every year millions of people throughout the world are diagnosed using $^{99m}$Tc. However, neutron rich Tc radionuclides, even  $^{99m}$Tc are not well suited because of extremities in terms of half lives for laboratory experiments of tens of hours duration. On the other hand proton rich $\gamma$ and $\beta^{+}$ emitting Tc tracers ($^{93}$Tc, $^{94m, 94}$Tc,$^{95m, 95}$Tc, $^{96}$Tc) exhibit required salient features to serve diverse applications, particularly in nuclear medicine. For example, $^{94m}$Tc (T$_{1/2}$=52.0 m, I$_{\beta^{+}}$=70.2\%, E$_{\beta^{+}}$ = 2.44 MeV) is potentially promising radionuclide for PET (Positron Emission Tomography) imaging of organs because of its high positron emission energy. Short lived $\gamma$-emitters $^{93}$Tc (T$_{1/2}$=2.75 h) and $^{94}$Tc (T$_{1/2}$=293 m) are convenient for short span experiments while $\gamma$-emitter $^{95}$Tc, which has comparatively long half life of 20 h, can be useful to keep track of the metabolic function of human brain and heart. Therefore, information on the nuclear data for the production of proton rich Tc tracers having considerable half life is important.

Several studies have already been reported which aimed to enquire the nuclear properties of proton rich Tc isotopes as well as isomers.
Usually proton rich tracers are produced in medium energy accelerators using light ion ($p$, $d$, $^{3}$He, $\alpha$) induced reactions  on molybdenum (Mo) or niobium (Nb) target \cite{graf, branquinho, ramamoorthy, denzler, Auler, fassbender, christian, strohmaier, tarkanyi,hogan1, hogan2, hogan3, hogan4, poppe, shakun, Izumo, lagunas, roesch93,uddin,uddinARI,khandaker,randa,bonardi,SL,kr}.  $^{93}$Nb is mononuclidic in nature. Therefore, investigation of nuclear properties of Tc isotopes, which are produced by $\alpha$ and $^{3}$He induced reactions on $^{93}$Nb, has got importance in basic nuclear physics \cite{graf, branquinho, ramamoorthy, denzler, Auler, fassbender, christian, strohmaier, tarkanyi}. Large number of reports are available on the production cross section of Tc radionuclides from proton induced reactions on enriched isotopes of Mo covering wide energy range \cite{hogan1, hogan2, hogan3, hogan4, poppe, shakun, Izumo, lagunas, roesch93, strohmaier}. Excitation functions of $^{93,93m,94,94m,95,95m,96}$Tc from proton and deuteron induced reactions on natural Mo have been reported in ref \cite{uddin,uddinARI,khandaker,randa,bonardi}. Moreover, cross sections of $^{95,96m,96}$Tc have also been measured by neutron induced reactions on enriched $^{96}$Ru as well as natural ruthenium target \cite{luo,luoprc}. Even production of $^{95,96}$Tc by proton and $\gamma$-ray induced reactions on $^{99}$Tc have been reported in \cite{zaitseva,sekine}.

However, cross section data of proton rich Tc isotopes produced via heavy ion induced reactions is rare. A few literatures are available where excitation functions of Tc isotopes are measured from $^{12}$C induced reactions on $^{93}$Nb and $^{89}$Y \cite{Misaelides, bindu} aiming to study the reaction mechanisms involved in the particular reaction. Unlike neutron and light charged particle reactions, understanding of reaction mechanisms as well as nuclear data involved in the heavy ion reactions are not adequate till date and hence demands much attention in this direction.

In order to satisfy the growing demand of nuclear data of clinically and biologically important proton rich radioisotopes of Tc, in this paper, we have investigated two separate routes, e.g., $^{7}$Li + $^{nat}$Zr and $^{9}$Be + $^{nat}$Y. Excitation functions of evaporation residues produced in $^{7}$Li+$^{nat}$Zr and $^{9}$Be +$^{nat}$Y have been measured in the incident energy range 37-45 MeV and 30-48 MeV respectively using standard stacked foil technique followed by off line $\gamma$-spectrometric studies. Other radionuclidic impurities are also produced in these two routes along with Tc. The word $\it {impurity}$ in the perspective of nuclear data useful in the field of nuclear medicine has been defined in our recent paper \cite{mm}. The impurity describes all the evaporation residues (stable as well as radionuclidic) and their decay products including the decay of compound nucleus when aimed for a particular radionuclide. Quantification as well as  reduction of those impurities by selecting proper nuclear reaction parameters like, incident energy, target thickness etc. are important. Cross section data provides idea of selecting those parameters. Enriched isotopes can reduce the production of impurities, but are quite expensive. Naturally occurring mononuclidic elements are preferred as target material to prevent opening of less reaction channels. However, it is also possible to use stable elements with more than one naturally occurring isotopes, provided the production parameters are controlled.

Therefore, in the present work, our goal is twofold: \\
(i) measurement of excitation functions of evaporation residues produced in $^{7}$Li+$^{nat}$Zr and $^{9}$Be+$^{nat}$Y reactions \\
(ii) investigation of the reaction mechanisms involved in those two reactions comparing with the nuclear reaction model predictions.

Broadly, nuclear reaction is understood in terms of three kinds of reaction mechanisms, namely, direct (DIR), preequilibrium (PEQ) and equilibrium or evaporation (EQ). DIR reaction is the single step process that occurs due to large momentum transfer from projectile to the target nucleons. This results in high energy ejectile emission leaving residual at lower excited states. Compound nucleus is formed when incoming projectile energy is shared between all the nucleons in the target nucleus and equilibrium emissions (EQ) occur due to statistical fluctuation in energy. Between these two extremities, PEQ reaction process plays a significant role having features of both. Cross section of the product nuclei corresponding to a particular reaction is the resultant of all the reaction mechanisms involved.
In the present work, contribution from DIR reaction processes is not expected because of low incident energies. We have tried to explain the experimental cross sections of evaporation residues in terms of PEQ and EQ reactions in the incident energy range 37-45 MeV and 30-48 MeV respectively, using nuclear reaction model codes PACE-II \cite{pace} and ALICE91 \cite{alice1, alice2}. 

Section II presents the discussions about the protocol and formalism of the code PACE-II and ALICE91 required for the theoretical calculation of excitation functions. Experimental procedure is described in section III and section IV deals with the results and discussion of the present work.

\section{Nuclear reaction model calculation}
\subsection{PACE-II}

The code PACE-II is the modified version of the Monte-Carlo code Projection Angular-momentum Coupling Evaporation. The deexcitation process of the excited nuclei is calculated using the modified version of the code JULIAN, which follows the correct procedure of angular momentum coupling at each stage of deexcitation using Hauser-Feshbach \cite{Hauser} model.
The transmission coefficients for light particle emission of neutrons, protons and $\alpha$s are determined from the optical model potential where all the optical model parameters are taken from ref \cite{perey}. The shift in coulomb barrier during deexcitation is accounted by calculating the transmission coefficients at an effective energy determined by the shift. The code internally decides level densities and masses it needs during deexcitation. Because of low excitation energy, Gilbert-Cameron level density prescription is used in the present work  with $a$, level density parameter equals to A/12 MeV$^{-1}$. The ratio of $a_{f}$/$a_{n}$ is chosen as unity. Fission is considered as a decay mode. The finite range fission barrier of Sierk \cite{sierk} has been used.  Compound nuclear fusion cross section is determined by using the Bass method \cite{bass}. Yrast parameter is taken as unity. A non-statistical yrast cascade $\gamma$-decay chain has been artificially incorporated to simulate $\gamma$ multiplicity and energy results. 

\subsection{ALICE91}

The code ALICE91 \cite{alice1,alice2} has been used to calculate the excitation function of product radionuclides. Geometry dependent hybrid model \cite{alice2,hyb1,hyb2,hyb3} has been used to calculate PEQ emissions and Weisskopf-Ewing formalism \cite{weisskopf} for EQ emissions. 
The hybrid model is the combination of exciton model \cite{exciton} and Boltzmann master equation approach \cite{bme1, bme2}. It assumes that the target-projectile composite system proceeds through two body interaction process. Each stage of the relaxation process is designated by the total number (n) of excited particles, i.e., sum of the excited particles (p) and holes (h). In each two body interaction, p-h pair may be created or annihilated or redistribution of energy takes place without changing the number. Hybrid model uses {\it never come back} approximation, i.e, the model assumes only p-h pair is created in each interaction. Hybrid model explicitly determines the pre-emission energy distribution of the excited particles which helps to estimate high energy emissions more accurately. Geometry dependent hybrid model includes the nuclear surface effects \cite{alice2, hyb2,hyb3}. The PEQ emission cross section for a particular ejectile $x$ with energy $\epsilon_{x}$ is given by

\begin{equation}
\label{mmeq3}
\begin{array}{lll}
\sigma _{PEQ}(\epsilon _{x}) & = 
 & \frac{\lambda ^{2}}{4\pi}\sum\limits_{l=0}^{\infty}(2l+1)T_{l}\sum  \limits
^{\overline{n}}_{n=n_{0},\atop {\Delta n=2}}D_{n}\left[f^{x}_{n}\frac{N_{n}(l,U,\epsilon_{x})}{N_{n}(l,E_{c})}\right]{\lambda _{c}(\epsilon _{x})\over{\lambda _{c}(\epsilon _{x})+\lambda _{t}(\epsilon _{x})}}
\end{array}
\end{equation}

Here, $\lambda$ is the de-Broglie wave length of the projectile, $T_{l}$ is the transmission coefficient of the $l^{th}$ partial wave, $D_{n}$ is the depletion factor of the $n^{th}$ exciton state, that is, probability of reaching $n$ exciton state without prior emission and $f^{x}_{n}$   is number of $x$ type excited nucleon present in it. The number $n_{0}$ and $\overline n$  are the initial and equilibrium exciton numbers respectively. The ratio ${N_{n}(l,U,\epsilon_{x})}/{N_{n}(l,E_{c})}$  is the probability of finding $x$ type nucleon in the $n$ exciton state with energy ($\epsilon_{x}+B_{x}$)  where $B_{x}$ is the separation energy of $x$.  The factor  ${\lambda _{c}(\epsilon 
_{x})\over{\lambda _{c}(\epsilon _{x})+\lambda _{t}(\epsilon _{x})}}$  is the emission probability of $x$ with energy $\epsilon_{x}$. $\lambda _{t}(\epsilon _{x})$  is the two-body interaction rate. The emission rate $\lambda _{c}(\epsilon _{x})$  is calculated by \cite{alice2}
\begin{equation}
\label{mmeq4}
\lambda_{c}(\epsilon _{x})=\frac{(2S_{x}+1)\mu_{x}\epsilon_{x}\sigma_{inv}(\epsilon_{x})}{\pi^{2}\hbar^{3}g} 
\end{equation} 
where, $S_{x}$  is the intrinsic spin of $x$, $\mu_{x}$   is the reduced mass, $\sigma_{inv}$  is the inverse cross section of the ejectile $x$ with energy $\epsilon_{x}$  being absorbed by the residual and $g$  is the single particle level density of the composite nucleus. 
EQ emission cross section is calculated using Weisskopf-Ewing formalism as
 \begin{equation}
\label{mmeq5}
\sigma_{EQ}(\epsilon _{x}) \sim \sigma_{comp}\frac{e^{2(aU)^{1/2}}}{U}
\end{equation}
$\sigma_{comp}$  is the compound nuclear formation cross section, $a$ is the level density parameter and $U$ is the available excitation energy of the compound nucleus after the PEQ emissions. $\sigma_{comp}$  is calculated as  $\sigma_{comp}=\sigma_{abs}-\sigma_{PEQ}$, where $\sigma_{abs}$  is the absorption cross section of the projectile in the target and $\sigma_{PEQ}$  is the total PEQ emission cross section. 
  
The calculations have been performed using the code ALICE91 \cite{alice1,alice2} with geometry dependent hybrid model for PEQ emissions and Weisskopf-Ewing formalism for EQ emissions. $n$, $p$, $d$ and $\alpha$ emissions are considered from the residual nuclides of 12 mass unit wide and 10 charge unit deep including the composite nucleus. Fermi gas level density has been used for the calculation of reaction cross sections. Reverse channel reaction cross sections have been calculated using the optical model. The level density parameter, $a$ is taken as A/12. Rotating finite range fission barriers of Sierk has been chosen. Total number of nucleons in the projectile has been chosen as the initial exciton number for the PEQ cross section calculation.

\section{Experimental procedure}
\subsection{Irradiation parameters}
\subsubsection{$^{7}$Li+$^{nat}$Zr}

Pure metallic Zr foil (99.94\%) was procured from the Johnson Matthey \& Co. Limited.  Natural assey of Zr contains $^{90}$Zr (51.45\%), $^{91}$Zr (11.22\%), $^{92}$Zr (17.15\%), $^{94}$Zr (17.38\%) and $^{96}$Zr (2.8\%). To measure the excitation function, self-supporting metallic Zr foils of about 3 mg/cm$^{2}$ thickness were prepared. Experiments were carried out to measure excitation functions of the product radionuclides at BARC-TIFR Pelletron at Mumbai, India. The stack of target-catcher assembly was prepared by placing a Zr foil followed by an aluminum catcher foils of thickness 1.5 mg/cm$^{2}$ and was bombarded by $^{7}$Li$^{3+}$ and total charge of 1425 $\mu$C was collected over 6.25 h duration. Excitation functions of evaporation residues of $^{93,94,95,96}$Tc, $^{93m}$Mo and $^{90,96}$Nb were measured in the projectile energy range 37-45 MeV. 

\subsubsection{$^{9}$Be+$^{nat}$Y}

The only naturally abundant isotope of yttrium is $^{89}$Y, which was procured from Alfa Aesar (99.9\% pure). Thin self supporting yttrium foils of thickness 3-3.8 mg/cm$^{2}$  were prepared to measure the excitation function. Target stack was assembled by placing aluminum catcher foils of thickness 1.5 mg/cm$^{2}$ in between Y foils. Target assembly was bombarded by $^{9}$Be$^{4+}$ beam and total charge of 304 $\mu$C was collected over 3.75 h duration. Excitation functions of $^{93,94,95}$Tc and $^{93m}$Mo were measured in the 30-48 MeV incident energy range. 

\subsection{Measurement of activity}
The aim of this paper was to study the production of $^{93,94,95,96}$Tc radionuclides and the associated radionuclidic impurities produced from both the reactions. The duration of the irradiation time was chosen according to the beam intensity and the half life of the product nuclides. Target and aluminum foils were mounted on aluminum ring of 10 mm inner and 22 mm outer diameter with 0.5 mm thickness. The residual products, if any, recoiled in the beam direction, were completely stopped in the aluminum backing. Large area of the catcher foils ensures the complete collection of recoiled evaporation residues. 
At the end of the bombardment, foils were counted for the $\gamma$-ray activity of the evaporation residues by an HPGe detector of 2.13 keV resolution at 1332 keV coupled with a PC based MCA, PCA2 (OXFORD). Efficiency calibration of the detector was performed as a function of $\gamma$-ray energy using a standard $^{152}$Eu (T$_{1/2}$=13.506 y) source of known activity. Each foil was counted for 300 seconds in the live time mode leaving proper cooling time after the bombardment and successive measurements were carried out for sufficiently long time in the same geometry.
In general, projectile energy at a target is the average of incident and outgoing beam energy. Beam energy degradation in the target and the catcher foils was calculated using the Stopping and Range of Ions in Matter (SRIM) \cite{srim}. Energy loss is less than  2\% in case of zirconium targets whereas it is about 4-5\% for yttrium targets. Total charge of each irradiation experiment was measured by an electron suppressed Faraday cup stationed at the rear of the target assembly which in turn is the measure of beam intensity. Evaporation residues recoiled to the catcher foils were also measured by $\gamma$-ray counting. It was noticed that minute amount of radionuclides recoiled to the catcher foil only at higher incident energies
and there by neglected in the present calculation.

The nuclear spectroscopic data of the radionuclides studied in the present work is enlisted in the Table \ref{mmt1} \cite{decaydata}. The product yields of the evaporation residues in each foil were calculated using the reported nuclear data \cite{decaydata}; half life, $\gamma$-energy, branching ratio, etc. Background subtracted peak area count correspond to a particular $\gamma$-ray energy of the $\gamma$-spectra is the measure of the yield. The product yield (y$_{i}$) of particular radionuclide (evaporation residue) $i$ at the end of bombardment was calculated from the standard relation

\begin{equation}
\label{mmeq1}
y^{i}=\frac{c(t)}{\epsilon^{i}_{\gamma}I^{i}_{\gamma}}e^{\lambda^{i} \tau} 
\end{equation} 
where $c(t)$ is the count rate at any time $t$, $\epsilon^{i}_{\gamma}$ and $I^{i}_{\gamma}$ are the detection efficiency and branching intensity of the characteristic $\gamma$-ray of the evaporation residue, designated by $i$, with decay constant $\lambda^{i}$, $\tau$ is the cooling time. Cross section of the $i^{th}$ evaporation residue ($\sigma^{i}(E)$) at an incident energy, $E$ is calculated from the activation equation
\begin{equation}
\label{mmeq2}
y^{i}=I_{p}\sigma^{i}(E)n_{tg}x_{tg} (1-e^{-\lambda_{i}T})
\end{equation} 
where $I_{p}$ is the intensity of the projectile, $n_{tg}$ and $x_{tg}$ are the number of target nuclei per unit volume and target thickness, respectively, $T$ is the duration of irradiation.

\subsection{Uncertainties in measurements}

The expected errors associated in the cross section measurement are as follows:\\
(i)	Maximum error in efficiency calibration of HPGe detector	$\approx$ 2\%	\\
(ii)	Maximum error in determining target thickness ($n_{tg}x_{tg}$) in atoms/cm$^{2}$ $\approx$  5\% \\
(iii)	 Systematic error in the beam current that propagated to the cross section data $\approx$  10-12\% \\		
(iv)	Uncertainty in the incident beam energy at the successive targets may occur due to the energy degradation in the aluminum catchers. According to ref \cite{wilken, kemmer}, the energy straggling is expected to be small even in case of lowest incident energy and hence was neglected in the present work. 

Apart from the above, error occurs in the cross section data due to counting statistics. The total associated error related to the cross section measurement was determined considering all the factors discussed and the data presented up to 95\% confidence level.

\begin{figure}
\begin{center}
\includegraphics[height=8.0cm]{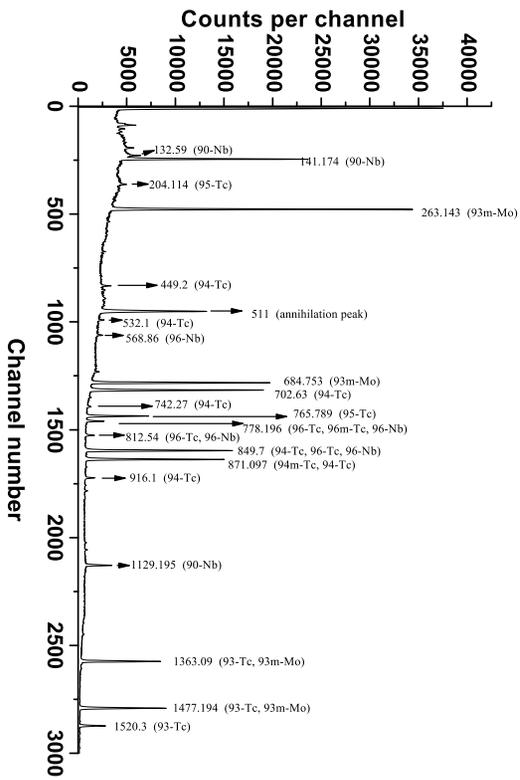}
\caption{$\gamma$-ray spectrum of the radionuclides produced in $^{7}$Li+$^{nat}$Zr reaction at 44.6 MeV incident energy} 
\label{fig1}
\end{center}
\end{figure}

\begin{figure}
\begin{center}
\includegraphics[height=8.0cm]{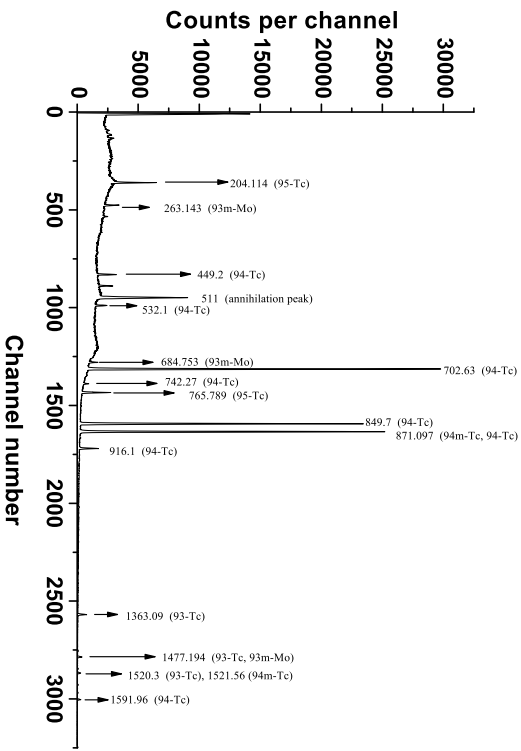}
\caption{$\gamma$-ray spectrum of the radionuclides produced in $^{9}$Be+$^{89}$Y reaction at 47.5 MeV incident energy} 
\label{fig2}
\end{center}
\end{figure}

\begin{table}
\caption{Nuclear spectrometric data of the radionuclides produced through $^{7}$Li+$^{nat}$Zr and $^{9}$Be+$^{89}$Y reactions. Bold $\gamma$-rays are used in determining the excitation functions.}
\label{mmt1}
\begin{tabular}{cccccccc}
\hline
    Product nuclei & Spin & T$_{1/2}$ & Decay mode (\%) & E$_{\gamma}$(keV) (I$_{\gamma}$\%)  \\
\hline
   
 	 $^{93}$Tc & 9/2$^{+}$ & 2.75 h & $\epsilon$(100) & {\bf1363.09} (66) \\
 	 $^{94m}$Tc & 2$^{+}$ & 52.0 m & $\epsilon$(100), IT($<$0.1) & 871.097(94.2)  \\
 	 $^{94}$Tc & 7$^{+}$ & 293 m & $\epsilon$ (100)& {\bf 702.63}(99.6),871.097(99.9)  \\
 	 $^{95}$Tc & 9/2$^{+}$ & 20.0 h & $\epsilon$(100) & {\bf765.789}(94.0), 1073.713(3.75)  \\
 	 $^{96m}$Tc & 4$^{+}$ & 51.5 m & IT (98), $\epsilon$(2) & 778.196(1.9), 1200.165 (1.09)  \\
 	 $^{96}$Tc & 7$^{+}$ & 4.28 d & $\epsilon$ (100) & 778.196 (99.76), {\bf812.54}(82)  \\
 	 $^{93m}$Mo & 21/2$^{+}$ & 6.87 h & IT(99.88),$\epsilon$(0.12) & 263.143(56.7), {\bf684.753}(99.7)\\
 	 $^{96}$Nb & 6$^{+}$ & 23.35 h & $\beta^{-}$ (100) & {\bf568.86}(56.8) \\
 	 $^{90}$Nb & 8$^{+}$ & 14.6 h & $\epsilon$ (100) & {\bf1129.195}(92.7) \\

\hline
\end{tabular}
\end{table}

\section{Results and discussion}

The residual radionuclides produced through various reaction channels of $^{7}$Li+$^{nat}$Zr and $^{9}$Be+Y reactions at the maximum incident energies of 44.6 MeV and 47.5 MeV, respectively are shown in Figs. \ref{fig1} and \ref{fig2} with their characteristic $\gamma$-rays. 
Experimental excitation functions of the residues are compared with the theoretical predictions of PACE-II and ALICE91 in Figs. \ref{fig3}-\ref{fig6}. Experimental cross sections are represented by various symbols with the associated error and theoretical predictions are shown by the lines (solid line for PACE-II and dashed line for ALICE91). Though the code ALICE91 takes care of the PEQ emissions, it has been observed that PEQ reaction has almost no contribution, except about 5\% in the highest incident energy, in producing the evaporation residues in the energy range studied in the present work.

\subsection{$^{7}$Li+$^{nat}$Zr}

\begin{figure}
\begin{center}
\includegraphics[height=8.0cm]{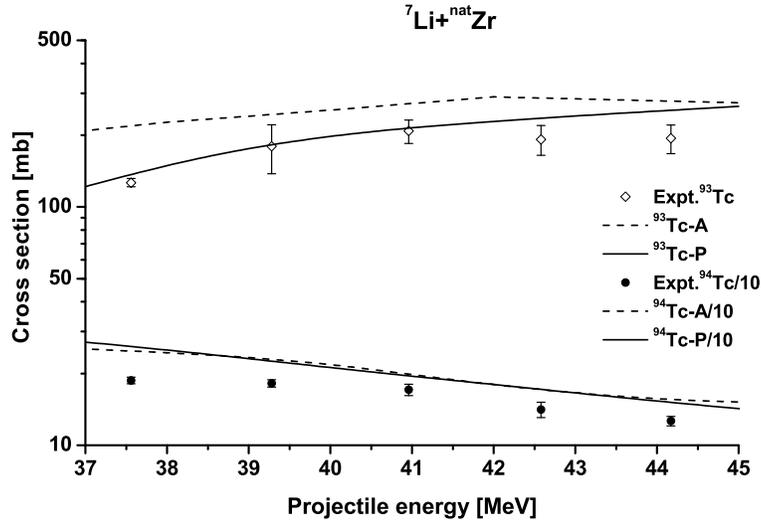}
\caption{Comparison between measured production cross sections of $^{93}$Tc and $^{94}$Tc with theoretical predictions of PACE-II and ALICE91. Cross section values are divided by 10 for $^{94}$Tc. } 
\label{fig3}
\end{center}
\end{figure}

\begin{figure}
\begin{center}
\includegraphics[height=8.0cm]{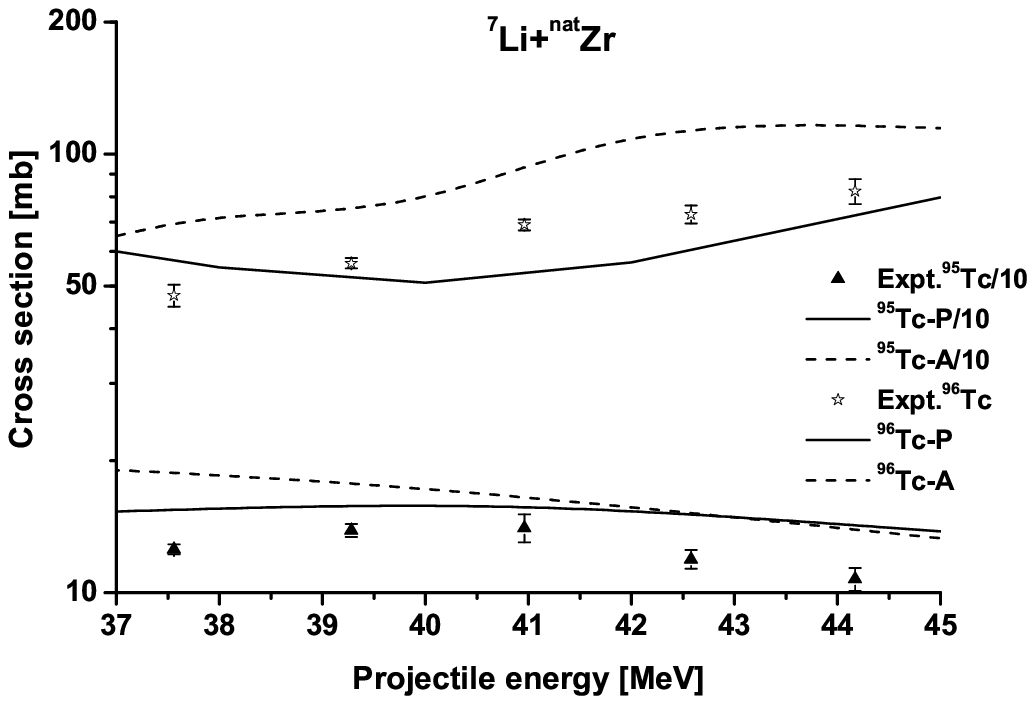}
\caption{Comparison between measured production cross sections of $^{95}$Tc and $^{96}$Tc with theoretical predictions of PACE-II and ALICE91. Cross section values are divided by 10 for $^{95}$Tc.} 
\label{fig4}
\end{center}
\end{figure}

\begin{figure}
\begin{center}
\includegraphics[height=8.0cm]{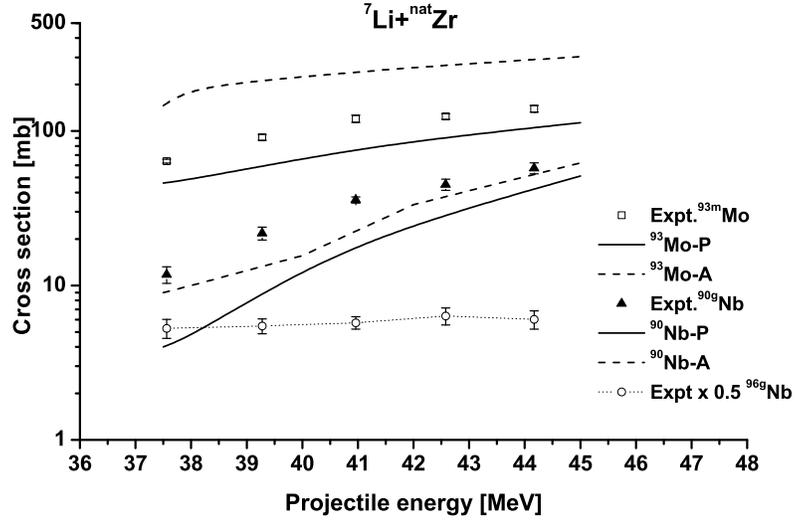}
\caption{Comparison between measured production cross sections of $^{93m}$Mo and $^{90,96}$Nb with theoretical predictions.  Cross section values are multiplied by 0.5 for $^{96}$Nb.} 
\label{fig5}
\end{center}
\end{figure}

Figure \ref{fig1} shows that Tc radionuclides, $^{93,94,95,96}$Tc, are produced along with $^{93m}$Mo and $^{90,96}$Nb due to the bombardment of 
$^{7}$Li on $^{nat}$Zr at 44.6 MeV incident energy. $^{93,94,95,96}$Tc radionuclides are produced through $^{nat}$Zr($^{7}$Li, $xn$), $x$ being the number of reaction channels. Measured excitation functions of $^{93,94}$Tc and $^{95,96}$Tc are compared with theoretical predictions in Figs. \ref{fig3} and \ref{fig4} respectively in the 37-45 MeV energy range.  $^{nat}$Zr contains five naturally abundant Zr isotopes; $^{90}$Zr (51.45\%), $^{91}$Zr (11.22\%) , $^{92}$Zr (17.15\%) , $^{94}$Zr (17.38\%) and $^{96}$Zr (2.8\%). The theoretical excitation functions reported in the Figs. \ref{fig3} - \ref{fig5} are calculated taking the weighted average of all five natural isotopes of Zr.

It is observed from Fig. \ref{fig3} that experimental cross sections of $^{93}$Tc are well reproduced by the prediction of PACE-II whereas ALICE91 overpredicts throughout the range with a maximum of 75\% at 37.5 MeV incident energy. In case of $^{94}$Tc, both the theoretical calculations are close to each other and slightly overpredict the experimental data. However, overall observation shows a good agreement between the experimental results and PACE-II predictions. 

Figure \ref{fig4} shows comparison between the measured cross sections of $^{95,96}$Tc  and the theoretical estimations. In both the cases, experimental data agree with the PACE-II calculations. However, ALICE91 overpredicts the data throughout the range. This observation reveals  that $^{93,94,95,96}$Tc radionuclides are produced as a result of complete fusion of $^{7}$Li in the $^{nat}$Zr target which is according to the expectation. It is observed from Figs. \ref{fig3} and \ref{fig4} that PACE-II calculations reproduced the measured excitation functions with a tendency of slight overprediction, which implies the prominent role of compound nuclear reaction mechanism.

Figure \ref{fig5} presents the experimental excitation function for the production of $^{93m}$Mo and $^{90,96}$Nb through $^{7}$Li+$^{nat}$Zr reactions along with the theoretical values. PACE-II underpredicts measured excitation functions of $^{93m}$Mo and $^{90}$Nb by 25\% and 50\% respectively, at lowest incident energy. ALICE91 calculation unexpectedly overpredicts the experimental excitation function of  $^{93m}$Mo by a factor of 2, whereas it agrees with the cross sections of $^{90}$Nb. Underprediction of cross section values by PACE-II might be an indication of the PEQ process in formation of $^{93m}$Mo and $^{90}$Nb. However, none of the theoretical models predicts the production of  $^{96}$Nb when its signature observed experimentally.

\subsection{$^{9}$Be+$^{89}$Y}

\begin{figure}
\begin{center}
\includegraphics[height=8.0cm]{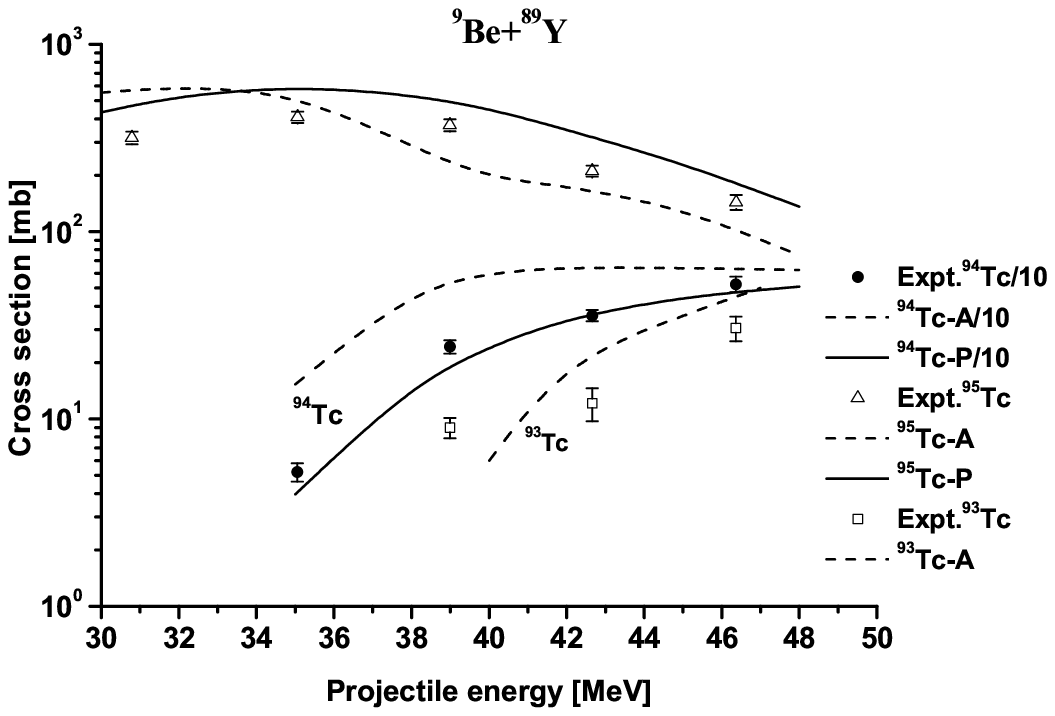}
\caption{Comparison between measured production cross sections of $^{95}$Tc, $^{94}$Tc and $^{93}$Tc with theoretical predictions of PACE-II and ALICE91. Cross section values are divided by 10 for $^{94}$Tc.} 
\label{fig6}
\end{center}
\end{figure}

Due to the bombardment of $^{9}$Be on $^{89}$Y, mainly $^{93,94,95}$Tc are produced in the 30-48 MeV incident energy range. Only a small amount of $^{93m}$Mo is produced at the 47.5 MeV incident energy. Figure \ref{fig6} shows the experimental excitation function of $^{93,94,95}$Tc along with the theoretical predictions of PACE-II and ALICE91.  Measured excitation functions of $^{94}$Tc and $^{95}$Tc are well evaluated by PACE-II calculations whereas ALICE91 overpredicts the measured cross section of $^{94}$Tc. The incident energy range reported (30-48 MeV) in the present work covers maximum production of $^{95}$Tc and $^{94}$Tc at the lowest and highest incident energy side. The trend of the measured excitation function exactly matches with the PACE-II predictions. 
Measured excitation function of $^{93}$Tc is compared only with the ALICE91 values, which starts from 40 MeV incident energy  and slightly overpredicts the experimental cross sections at next two energy values. However, cross section has been measured at 39 MeV, which is well above the threshold value, 36.5 MeV. According to PACE-II, $^{89}$Y($^{9}$Be,5$n$) reaction channel will open from 45 MeV.  It is clear from the cross section values that Tc radionuclides are produced  through $^{89}$Y($^{9}$Be, $xn$), ($n$=3,4,5), reaction channels by complete nuclear fusion. This is expected when the maximum incident energy is less than 5.5 MeV/nucleon. No signature of PEQ process has been observed from the experimental cross sections. Production of $^{93m}$Mo is observed only in the first foil at 47.5 MeV. The measured cross section is (13 $\pm$ 2.1) mb, which was expected from both the theoretical estimations.

The basic difference in the two theoretical predictions from PACE-II and ALICE91 is observed due the formalisms adopted for the simulation of compound nuclear processes. ALICE91 uses faster approach of Weisskopf-Ewing sacrificing rigor in physics whereas PACE-II takes Hauser-Feshbach formalism sacrificing computing time. The later one found to be better in explaining the nuclear data obtained entirely from the compound nuclear reactions. The present work agrees with this fact.

\section{Conclusion}

Present work reports for the first time two production routes of  proton rich Tc radionuclides via heavy ion induced reactions of $^{7}$Li+$^{nat}$Zr and $^{9}$Be+$^{89}$Y. Excitation functions of the evaporation residues produced from those reactions have been measured in energy range 37-45 MeV and 30-48 MeV respectively. The measured cross section values were found to be in good agreement with the detail Hauser-Feshbach model calculations using PACE-II.  Cross section data essentially revealed the compound nuclear reaction mechanism as per expectation in the energy range studied. Experimental nuclear reaction data is important for the quality production of Tc radionuclides.

In case of $^{7}$Li+$^{nat}$Zr reaction,  reported incident energy range, 37-45 MeV, is not sufficient to get a complete picture of excitation functions of evaporation residues. More experimental data is needed at both, higher as well as lower incident energies. 
However, presence of all other radionuclides, which may cause difficulty in optimizing production parameters of a particular isotope of interest reducing other radionuclidic impurity, is prominent from the  study.
The overall knowledge of the work will be helpful in producing clinically important proton rich Tc radionuclides. Cross section data will also help to increase radionuclidic purity optimizing nuclear reaction parameters. However, it is once again clear that mononuclidic target has advantage in this context.

\begin{acknowledgments}

This work has been carried out as a part of  the Saha Institute of Nuclear Physics-Department of Atomic Energy, XI five year plan project "Trace Analysis: Detection, Dynamics and Speciation (TADDS)". Authors are thankful to A.G. Mahadkar from TIFR and Pradipta Das from SINP for preparing zirconium and yttrium targets. Thanks to pelletron staff of TIFR pelletron for their kind support and help during the experiments. 

\end{acknowledgments}

\end{document}